\documentclass[twocolumn,showpacs,aps,pra,superscriptaddress]{revtex4-1}

\usepackage{ulem,bm}
\usepackage[colorlinks, linkcolor=blue, citecolor=blue, urlcolor=blue]{hyperref}
\usepackage{array,epsfig,amsmath,amssymb}
\usepackage{graphicx} 
\usepackage{dsfont}
\usepackage{color}

\def\bra#1{\langle #1|}
\def\ket#1{|#1 \rangle}
\def\bracket#1#2{\langle #1|#2 \rangle}

\begin{document}

\title{Nearly deterministic Bell measurement with multiphoton entanglement for efficient quantum information processing}

\author{Seung-Woo Lee}
\affiliation{Center for Macroscopic Quantum Control, Department of Physics and Astronomy, Seoul National University, Seoul, 151-742, Korea}
\author{Kimin Park}
\affiliation{Center for Macroscopic Quantum Control, Department of Physics and Astronomy, Seoul National University, Seoul, 151-742, Korea}
\affiliation{Department of Optics, Palack\'y University, 17. listopadu 1192/12, 77146 Olomouc, Czech Republic}
\author{Timothy C. Ralph}
\affiliation{Centre for Quantum Computation and Communication Technology, School of Mathematics and Physics, University of Queensland, St Lucia, Queensland 4072, Australia}
\author{Hyunseok Jeong}
\affiliation{Center for Macroscopic Quantum Control, Department of Physics and Astronomy, Seoul National University, Seoul, 151-742, Korea}

\begin{abstract}
We present a detailed analysis of the Bell measurement scheme proposed in Lee {\em et al.} [Phys. Rev. Lett. {\bf 114}, 113603 (2015)] based on a logical qubit using Greenberger-Horne-Zeilinger entanglement of photons. The success probability of the proposed Bell measurement can be made arbitrarily high using only linear optics as the number of photons in a logical qubit increases. We compare our scheme with all the other proposals, using single-photon qubits, coherent-state qubits or hybrid qubits, suggested to enhance the efficiency of the Bell measurement. As a remarkable advantage, our scheme requires only photon on-off measurements, while photon number resolving detectors are necessary for all the other proposals. We find that the scheme based on coherent-state qubits shows the best performance with respect to the attained success probability in terms of the average number of photons used in the process, while our scheme outperforms the schemes using single-photon qubits. We finally show that efficient quantum communication and fault-tolerant quantum computation can be realized using our approach. 
\end{abstract}

\maketitle

\section{introduction}
\label{sec:intro}

Bell measurement is a crucial element of quantum communication and quantum computation protocols. It discriminates between four entangled states known as Bell states, 
\begin{equation}
\label{eq:BellStates}
\begin{aligned}
\ket{\Phi^\pm}=\frac{1}{\sqrt{2}}(\ket{0_L}\ket{0_L}\pm\ket{1_L}\ket{1_L}),\\
\ket{\Psi^\pm}=\frac{1}{\sqrt{2}}(\ket{0_L}\ket{1_L}\pm\ket{1_L}\ket{0_L}),
\end{aligned}
\end{equation}
where $\ket{0_L}$ and $\ket{1_L}$ are the logical qubit bases. In optical implementations, single photons with their polarization degree of freedom are typically employed to construct logical qubits \cite{PKok2007,Ralph2010,Knill2001}. For a single photon qubit, the Bell measurement can be implemented using linear optics and photodetectors \cite{Lut99,Calsa2001}, which we shall refer to as the standard Bell-measurement scheme. 
This scheme, in effect, projects two photons onto a complete measurement basis of two Bell states and two product states so that only two of the Bell states can be unambiguously identified.
Due to this reason, the success probability of the standard Bell measurement is limited to 50\% \cite{Lut99,Calsa2001}, and it has been a fundamental hindrance to the implementation of a deterministic quantum teleportation and scalable quantum computation with linear optics \cite{PKok2007,Ralph2010}.

In order to enhance the success probability of the Bell measurement, several schemes have been proposed based on single-photon qubits using either ancillary entangled states \cite{Grice2011,Ewert2014} or additional squeezing operations \cite{Zaidi2013}. Instead of using single photons, different types of physical degrees of freedom of light can also be employed to construct logical qubits, for example, using superpositions of coherent states \cite{Jeong2001,Jeong2002} or hybrid entanglement of single photons and coherent states  \cite{SLee13,SWLEE13}. Although each of these schemes has its own merit, there exist several features to overcome in the implementation of those schemes, such as the requirement of ancillary resource entanglement \cite{Grice2011,Ewert2014}  or the limited success probabilities \cite{Zaidi2013}. Moreover, all aforementioned schemes suffer from the requirement of photon number resolving detection 
\cite{Grice2011,Zaidi2013,Ewert2014,Jeong2001,Jeong2002,SWLEE13}.

In this paper, we discuss a scheme for implementing a nearly deterministic Bell measurement with linear optics and photon on-off measurements \cite{SLee15}. The logical qubit is defined as a Greenberger-Horne-Zeilinger (GHZ) type multiphoton entangled state. It is shown that the logical Bell states can be efficiently discriminated by performing $N$ times of standard Bell measurements, where $N$ is the number of photons in a logical qubit. 
The limited success probability of the standard Bell measurement 1/2 is overcome by the fact that each of the four logical Bell states is characterized by the number of contributions from two Bell states of single photon qubits, that can be identified by the standard Bell measurement performed on photon pairs. As a result, the logical Bell measurement fails only when none of the $N$ pairs is a detectable Bell state, resulting in a success probability of $1-2^{-N}$ that rapidly approaches unity as $N$ increases. A detailed review of our scheme will be presented in Sec.~\ref{sec:BM}.

In Sec.~\ref{sec:compare}, we compare our scheme with other well known Bell measurement schemes proposed based on linear optics; our scheme outperforms the previous proposals employing single photon qubits \cite{Grice2011,Zaidi2013,Ewert2014} in its efficiency against the number of photons used in the measurement process. The other schemes employing coherent state \cite{Jeong2001,Jeong2002}  or optical hybrid qubits \cite{SLee13,SWLEE13} enable us to achieve even higher success probability than ours with the same average photon number usage. However, all the other schemes except ours require photon number resolving detection, while our scheme can be performed with linear optics and photon on-off measurements only. 

In Sec.~\ref{sec:App}, we discuss the applications of our scheme to quantum communication and computations. We will show that a qubit in an $N$ photon GHZ-type entanglement can be teleported with an arbitrarily high success probability with a $2N$ photon GHZ-type entangled channel as $N$ becomes large. It will be shown that in our approach a complete set of universal gate operations can be constructed using only linear optics, on-off measurements and multiphoton entanglement. We investigate the effect of photon losses, a major factor of errors, and demonstrate a fault-tolerant quantum computation in our approach. We expect our scheme to be a competitive approach to photonic quantum information processing.

\section{Linear optical Bell measurement scheme with multiphoton encoding}
\label{sec:BM}

\subsection{Standard Bell measurement scheme}

We first consider the standard Bell-measurement technique for single photon qubits. The logical basis of single photon qubits are typically defined with the photon polarization degree of freedom, in terms of either the horizontal and vertical polarization states, i.e. $\{\ket{H}, \ket{V}\}$, or the diagonal states $\ket{\pm}=(\ket{H}\pm\ket{V})/\sqrt{2}$. Here we assume the Bell states of single photon qubits in the diagonal basis,
\begin{equation}
\label{eq:BellStatesS}
\begin{aligned}
\ket{\Phi^\pm}=\frac{1}{\sqrt{2}}(\ket{+}\ket{+}\pm\ket{-}\ket{-}),\\
\ket{\Psi^\pm}=\frac{1}{\sqrt{2}}(\ket{+}\ket{-}\pm\ket{-}\ket{+}).
\end{aligned}
\end{equation}

The standard Bell-measurement scheme for single photon qubits employs linear optic elements such as polarizing beam splitter (PBS), wave plates and photon detection \cite{Calsa2001}. It projects two photons onto a complete measurement basis of two Bell states and two product states so that only two of the Bell states can be identified. For example, one can identify $\ket{\Phi^{-}}$ and $\ket{\Psi^{-}}$ using three PBS and four on-off photodetectors as illustrated in Fig.\ref{fig:scheme}(a). For $\ket{\Phi^-}$ or $\ket{\Psi^-}$ states, two photons are separated into different modes at the first PBS, resulting in one click from the upper two and another from lower two detectors. If the separated clicks have the different polarization output, i.e., $(H,V)$ or $(V,H)$, the result is $\ket{\Phi^{-}}$, while for the same polarization output, i.e., $(H,H)$ or $(V,V)$, the result is $\ket{\Psi^{-}}$. On the other hand, two photons of the state $\ket{\Phi^+}$ or $\ket{\Psi^+}$ proceed together via the first PBS either way to the upper or lower two detectors. All possible clicks at the detectors from $\ket{\Phi^+}$ can equally be yielded from $\ket{\Psi^+}$ so that it is impossible to discriminate them. Therefore, the success probability of the standard Bell measurement scheme is limited to 1/2 \cite{Lut99,Calsa2001}. Which two Bell states are successfully identified can be selected by putting appropriate wave plates at the input modes of the first PBS. We will refer to the standard Bell measurement in Fig.\ref{fig:scheme}(a) as $\rm B_s$ hereafter.

\subsection{Our proposal: Bell measurement with multiphoton encoding}

Here we introduce a novel scheme for discriminating four Bell states by using linear optics and qubits with multiphoton encoding \cite{SLee15}. We define the logical basis of qubits with $N$ photons as
\begin{equation}
\label{eq:Nencoding}
\begin{aligned}
& \ket{0_{L}}\equiv\ket{+}^{\otimes N}=\ket{+}_1\ket{+}_2\ket{+}_3 \cdot\cdot\cdot \ket{+}_N,\\
& \ket{1_{L}}\equiv\ket{-}^{\otimes N}=\ket{-}_1\ket{-}_2\ket{-}_3 \cdot\cdot\cdot \ket{-}_N.
 \end{aligned}
\end{equation}
Thus, a logical qubit is generally written in the form of a GHZ-type state as $\alpha |+\rangle^{\otimes N} + \beta |-\rangle^{\otimes N}$. The encoding of a logical qubit into $N$ photons can be realized, for example, by the teleportation scheme between an encoded single photon and $N$ photons. For this purpose, the GHZ state of $N+1$ photons is used as the teleportation channel and the encoded single photon as the input state.

\begin{figure}[tp]
\centering
\includegraphics[width=3.4in]{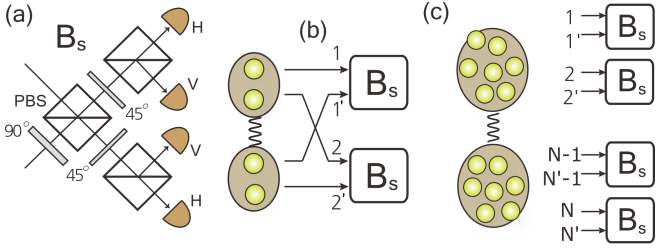}
\caption{(Color online). 
(a) Standard Bell measurement, $\rm B_s$, for single photon qubits using linear optical elements and photon on-off detectors. (b) Logical Bell measurement for the qubits encoded with two photons. (c) Logical Bell measurement for $N$-photon qubits through $N$ times of $\rm B_s$ measurements.}\label{fig:scheme}
\end{figure}

Let us first explain our proposal with the simplest case, i.e., two-photon encoding ($N=2$), where the logical qubit basis are $\ket{0_L}\equiv\ket{+}\otimes\ket{+}$ and $\ket{1_L}\equiv\ket{-}\otimes\ket{-}$. The logical Bell states in this encoding are written by 
\begin{equation}
\begin{aligned}
&\ket{\Phi^{\pm}_{(2)}}=\frac{1}{\sqrt{2}}(\ket{+}_{1}\ket{+}_{2}\ket{+}_{1'}\ket{+}_{2'}
\pm\ket{-}_{1}\ket{-}_{2}\ket{-}_{1'}\ket{-}_{2'}),\\
&\ket{\Psi^{\pm}_{(2)}}=\frac{1}{\sqrt{2}}(\ket{+}_{1}\ket{+}_{2}\ket{-}_{1'}\ket{-}_{2'}
\pm\ket{-}_{1}\ket{-}_{2}\ket{+}_{1'}\ket{+}_{2'}),
\end{aligned}
\end{equation}
where the first logical qubit is of photonic modes $1$ and $2$ while the second is of $1'$ and $2'$. If we simply rearrange the order of photon modes $1'$ and $2$, and by using the fact that $\ket{\pm}\ket{\pm}=(\ket{\Phi^{+}}\pm\ket{\Phi^{-}})/\sqrt{2}$ and $\ket{\pm}\ket{\mp}=(\ket{\Psi^{+}}\pm\ket{\Psi^{-}})/\sqrt{2}$, these Bell states can be represented in terms of the Bell states in Eq.~(\ref{eq:BellStatesS}) as
\begin{equation}
\begin{aligned}
&\ket{\Phi^{\pm}_{(2)}}=\frac{1}{\sqrt{2}}(\ket{\Phi^+}_{11'}\ket{\Phi^{\pm}}_{22'}+\ket{\Phi^-}_{11'}\ket{\Phi^{\mp}}_{22'}),\\
&\ket{\Psi^{\pm}_{(2)}}=\frac{1}{\sqrt{2}}(\ket{\Psi^+}_{11'}\ket{\Psi^{\pm}}_{22'}+\ket{\Psi^-}_{11'}\ket{\Psi^{\mp}}_{22'}).
\end{aligned}
\end{equation}
Interestingly, these four Bell states can be thus identified by means of two independent standard Bell measurements for single photon qubits ($\rm B_s$) performed on two photons, one is from the first and the other from the second qubit as illustrated in Fig.\ref{fig:scheme}(b). Note that the photons in each qubit are indistinguishable so that any single photon can be selected arbitrarily to perform each independent $\rm B_s$ measurement. 

Let us now explain how one can discriminate the four logical Bell states, keeping in mind that the $\rm B_s$ measurement defined in the previous section can unambiguously identify $\ket{\Phi^{-}}$ and $\ket{\Psi^{-}}$ out of the four Bell states of single photons. From the results of two $\rm B_s$ measurements, one can identify the Bell states encoded with two photons as follows: 
\begin{equation}
\nonumber
\begin{aligned}
&\ket{\Phi^{+}_{(2)}}:~{\rm both~B_s~succeed~with~}\ket{\Phi^-},\\
&\ket{\Phi^{-}_{(2)}}:~{\rm one~B_s~succeeds~with~}\ket{\Phi^-},\\
&\ket{\Psi^{+}_{(2)}}:~{\rm both~B_s~succeed~with~}\ket{\Psi^{-}},\\
&\ket{\Psi^{-}_{(2)}}:~{\rm one~B_s~succeed~with~}\ket{\Psi^{-}},\\
&{\rm Failure~}:~{\rm both~B_s~fail}.
\end{aligned}
\end{equation}
As the failure occurs only when both $\rm B_s$ measurements fail (i.e., neither $\ket{\Phi^{-}}$ nor $\ket{\Psi^{-}}$ is obtained), the four Bell states $\ket{\Phi^{\pm}_{(2)}}$ and $\ket{\Psi^{\pm}_{(2)}}$ can be discriminated with 75\% success probability, i.e., $P_s=3/4$. Note that our scheme employs only linear optics and on-off photodetectors as it can be implemented by means of two separate $\rm B_s$ measurements.

Our scheme can be generalized for the case of arbitrary $N$ photon encoding. From Eq.~(\ref{eq:Nencoding}), the logical Bell states $\ket{\Phi^{\pm}_{(N)}}=(|0_L\rangle|0_L\rangle\pm|1_L\rangle|1_L\rangle)/\sqrt{2}$ and $\ket{\Psi^{\pm}_{(N)}}=(|0_L\rangle|1_L\rangle\pm|1_L\rangle|0_L\rangle)/\sqrt{2}$ encoded with $N$ photons can be written by
\begin{equation}
\begin{aligned}
&\ket{\Phi^{\pm}_{(N)}}=\frac{1}{\sqrt{2}}\Big((\ket{+}\ket{+})^{\otimes N}\pm(\ket{-}\ket{-})^{\otimes N}\Big)\\
&=\frac{1}{\sqrt{2^{N+1}}}\Big((\ket{\Phi^{+}}+\ket{\Phi^{-}})^{\otimes N}\pm(\ket{\Phi^{+}}-\ket{\Phi^{-}})^{\otimes N}\Big),\\
&\ket{\Psi^{\pm}_{(N)}}=\frac{1}{\sqrt{2}}\Big((\ket{+}\ket{-})^{\otimes N}\pm(\ket{-}\ket{+})^{\otimes N}\Big)\\
&=\frac{1}{\sqrt{2^{N+1}}}\Big((\ket{\Psi^{+}}+\ket{\Psi^{-}})^{\otimes N}\pm(\ket{\Psi^{+}}-\ket{\Psi^{-}})^{\otimes N}\Big),
\end{aligned}
\end{equation}
by using $\ket{\pm}\ket{\pm}=(\ket{\Phi^{+}}\pm\ket{\Phi^{-}})/\sqrt{2}$ and $\ket{\pm}\ket{\mp}=(\ket{\Psi^{+}}\pm\ket{\Psi^{-}})/\sqrt{2}$. After some calculations, we can rewrite these in the form of 
\begin{equation}
\begin{aligned}
&\ket{\Phi^{+}_{(N)}}=\frac{1}{\sqrt{2^{N-1}}}\sum^{[N/2]}_{j=0}{\cal P}[\ket{\Phi^+}^{\otimes N-2j}\ket{\Phi^-}^{\otimes 2j}],\\
&\ket{\Phi^{-}_{(N)}}=\frac{1}{\sqrt{2^{N-1}}}\sum^{[(N-1)/2]}_{j=0}{\cal P}[\ket{\Phi^+}^{\otimes N-2j-1}\ket{\Phi^-}^{\otimes 2j+1}],\\
&\ket{\Psi^{+}_{(N)}}=\frac{1}{\sqrt{2^{N-1}}}\sum^{[N/2]}_{j=0}{\cal P}[\ket{\Psi^+}^{\otimes N-2j}\ket{\Psi^-}^{\otimes 2j}],\\
&\ket{\Psi^{-}_{(N)}}=\frac{1}{\sqrt{2^{N-1}}}\sum^{[(N-1)/2]}_{j=0}{\cal P}[\ket{\Psi^+}^{\otimes N-2j-1}\ket{\Psi^-}^{\otimes 2j+1}],
\end{aligned}
\end{equation}
where $[x]$ denotes the largest integer less than or equal to $x$ (i.e., $\leq x$), and ${\cal P}[\cdot]$ is the permutation function for $N$ elements of photon pairs. For example, we can rewrite the logical Bell state for $N=3$ as
\begin{equation}
\begin{aligned}
\ket{\Phi^{+}_{(3)}}&=\frac{1}{4}\Big((\ket{\Phi^{+}}+\ket{\Phi^{-}})^{\otimes 3}+(\ket{\Phi^{+}}-\ket{\Phi^{-}})^{\otimes 3})\Big)\\
&=\frac{1}{2}\Big(\ket{\Phi^{+}}\ket{\Phi^{+}}\ket{\Phi^{+}}+\ket{\Phi^{+}}\ket{\Phi^{-}}\ket{\Phi^{-}}\\
&~~~~~~~~+\ket{\Phi^{-}}\ket{\Phi^{+}}\ket{\Phi^{-}}+\ket{\Phi^{-}}\ket{\Phi^{-}}\ket{\Phi^{+}})\Big)\\
&=\frac{1}{2}\Big(\ket{\Phi^{+}}^{\otimes3}+{\cal P}[\ket{\Phi^{+}}\ket{\Phi^{-}}^{\otimes2}]\Big),
\end{aligned}
\end{equation}
where ${\cal P}[\ket{\Phi^{+}}\ket{\Phi^{-}}^{\otimes2}] = \ket{\Phi^{+}}\ket{\Phi^{-}}^{\otimes2}+\ket{\Phi^{-}}\ket{\Phi^{+}}\ket{\Phi^{-}}+\ket{\Phi^{-}}^{\otimes2}\ket{\Phi^{+}}$. Likewise, all other Bell states for $N=3$ can be represented in this way.

Therefore, the four logical Bell states with arbitrary $N$ photon encoding can be discriminated via $N$ times of separated $\rm B_s$ measurements as illustrated in Fig.\ref{fig:scheme}(c). Here, each $\rm B_s$ is performed on two photons; one is arbitrarily selected from the first and the other is from the second qubit. One can identify the logical Bell states from the results of $N$ independent $\rm B_s$ measurements as follows:
\begin{equation}
\nonumber
\begin{aligned}
&\ket{\Phi^{+}_{(N)}}:~{\rm even~number~of~B_s~succeed~with~}\ket{\Phi^-},\\
&\ket{\Phi^{-}_{(N)}}:~{\rm odd~number~of~B_s~succeed~with~}\ket{\Phi^-},\\
&\ket{\Psi^{+}_{(N)}}:~{\rm even~number~of~B_s~succeed~with~}\ket{\Psi^{-}},\\
&\ket{\Psi^{-}_{(N)}}:~{\rm odd~number~of~B_s~succeed~with~}\ket{\Psi^{-}},\\
&{\rm Failure~}:~{\rm all~B_s~fail}.
\end{aligned}
\end{equation}
As the failure occurs only when none of the $N$ times $\rm B_s$ measurements succeeds, the four logical Bell states $\ket{\Phi^{\pm}_{(N)}}$ and $\ket{\Psi^{\pm}_{(N)}}$ can be discriminated with a success probability $P_s=1-2^{-N}$. It is remarkable that the success probability rapidly approaches unity as $N$ increases. For example, the success rate up to 99.6 \% can be achieved with $N=8$ encoding. We note that our scheme for $N$ photon encoding can be performed effectively via either spatially or temporally distributed $N$ times of $\rm B_s$ measurements, irrespectively of the order of measurements. 
We will compare the efficiency of our scheme with other proposals in the following section.

\section{comparison with other proposals}
\label{sec:compare}

\subsection{Other schemes to beat the 1/2 limit}

Besides our proposal, many efforts have been devoted so far to improve the success probability of Bell measurement based on linear optics. We here consider several recent proposals employing single photon qubits \cite{Grice2011,Zaidi2013,Ewert2014}, and calculate its success probability in terms of the total number of photons ($\bar{n}$) used in the process for the comparison with our scheme. We will take into account the photons contained in the constructed logical qubits to be measured and in the ancillary states used in the measurement process as well as the increase of the average photon number by squeezing operation used in the process.

{\em Grice's scheme}-- A scheme proposed by Grice \cite{Grice2011} enhances the success probability of Bell-state discrimination employing ancillary entangled states of photons and photon number resolving detection in the Bell measurement process (see Ref.~\cite{Grice2011} for details). It increases the success probability up to $P_s=1-2^{-N_a}$ by using $2^{N_a}-2$ additional photons. Here $N_a$ is the number of ancillary entangled states $|\Gamma_j\rangle$ with $j=1,...,N_a-1$, each of which contains $2^{j}$ photons. The total number of photon used in the Bell measurement process can be thus obtained by counting all photons in two logical qubits and the ancillary states. As we here consider single photon encoding, two photons are used in two logical qubits. Thus, the total number of photons counted is $\bar{n}=2+2^{N_a}-2=2^{N_a}$, by which we can rewrite the success probability as $P_s=1-1/\bar{n}$. 

{\em Zaidi and van Loock's scheme} -- Zaidi and van Loock proposed a scheme to improve the success probability of Bell measurement for single photons in terms of inline squeezing operations accompanied by photon number resolving detections \cite{Zaidi2013}. With this approach, the success probability up to 64.3 \% is achieved for dual-rail encoding (photon polarization encoding) and to 62.5 \% for single rail encoding (vacuum and single photon state encoding). Let us here consider this scheme with dual-rail encoding in order to compare it with other schemes using single photon polarization. In this scheme, each mode of Bell states is squeezed by a squeezing operator $\hat{S}(r)=\exp[-r(\hat{a}^{\dagger 2}-\hat{a}^{2})/2]$ to improve the success probability of Bell discrimination. Here, $r$ is the squeezing parameter and $\hat{a}^{\dag}(\hat{a})$ is the creation (annihilation) operator. Suppose that Bell states, here assumed to be as $\ket{\phi^{\pm}}=(\ket{HH}\pm\ket{VV})/\sqrt{2}$, pass through a beam splitter as $U_\mathrm{BS}\ket{\phi^{\pm}}$ and then are squeezed. They can be written in the dual-rail representation as
\begin{equation}
\frac{i}{2}(\ket{2'0'0'0'}+\ket{0'0'0'2'}\pm\ket{0'2'0'0'}\pm\ket{0'0'2'0'})
\end{equation}
where $\ket{n'}=\hat{S}(r)\ket{n}$. 
The average photon number $\bar{n}$ in all four modes can be calculated by
\begin{equation}
\bra{\phi^{\pm}}\hat{S}_{1,2,3,4}(-r)(\hat{n}_1+\hat{n}_2+\hat{n}_3+\hat{n}_4)\hat{S}_{1,2,3,4}(r)\ket{\phi^{\pm}}
\end{equation}
where $\hat{n}_i$ is the photon number operator in the $i$-th mode and $\hat{S}_{1,2,3,4}(r)=\hat{S}_1(r)\hat{S}_2(r)\hat{S}_3(r)\hat{S}_4(r)$. If we use the relation
\begin{equation}
\hat{S}(-r)\hat{n}\hat{S}(r)=(\hat{a}^\dagger \cosh r- \hat{a} \sinh r)(\hat{a} \cosh r- \hat{a}^\dagger \sinh r),
\end{equation} 
the total average photon number in all four modes can be obtained as
\begin{equation}
\bar{n}
=\langle \hat{n}_1+\hat{n}_2+\hat{n}_3+\hat{n}_4\rangle=2\cosh 2r+4\sinh^2 r.
\end{equation}
The best value of $r$ to achieve the highest success probability is investigated in Ref.~\cite{Zaidi2013} and given as $r=0.6585$. With this value, the average photon number in the Bell state after the squeezing is obtained as $\bar{n}=6.00029$, which yields the success probability of the Bell measurement as $P_s=0.643$.  

{\em Ewert and van Loock's scheme 1}-- Ewert and van Loock proposed a scheme to discriminate the Bell states of single photon qubits by using ancillary multi-photon entanglement \cite{Ewert2014}. This employs a similar method used in Grice's scheme~\cite{Grice2011} and photon number resolving detection to increase the success probability. In the proposed Bell measurement setup, the total number of photons used in a single process can be counted as $\bar{n}=4N_m+2$ where $N_m$ is the number of ancillary states. The success probability achieved using $N_m$ ancillary entangled states is given by $P_s=1-2^{-{N_m}-1}$. Thus, we can rewrite the success probability by the average total photon number used in the process $\bar{n}$ as $P_s=1-2^{-{\bar{n}/4}-1/2}$.

{\em Ewert and van Loock's scheme 2}-- Recently, another scheme of linear optical Bell measurement was proposed \cite{Ewert15} by employing parity state encoding with multiphotons \cite{Ralph2005}. The logical qubit is constructed with $n$ blocks each containing $m$ photons. It was shown that the success probability reaches up to $P_s=1-2^{-n}$ similarly with our scheme. As the number of photons contained in each logical Bell state is $2\times n \times m$, we can write the success probability of Bell measurement by using the average photon number used in the process $\bar{n}=2nm$ as $P_s=1-2^{-\bar{n}/2m}$.

\subsection{Bell-measurement with other physical qubits}

Instead of single photons, other optical degrees of freedom can be used to construct the logical qubit in linear optical quantum information processing. We will here consider two previous proposals employing other optical qubits to compare with our scheme.

{\em Coherent state qubits}-- One of the well known approaches employs two coherent states $\ket{\alpha}$ and $\ket{-\alpha}$ with amplitudes $\pm\alpha$, or alternatively their superposed states $\ket{\alpha}\pm\ket{-\alpha}$, as a qubit basis. It was shown that nearly deterministic Bell-state measurement is possible in this approach \cite{Jeong2001,Jeong2002}. Quantum teleportation \cite{vanEnkPRA2001} and quantum computation schemes \cite{Jeong2001,Jeong2002,Ralph2003,Lund2008} based on the coherent state encoding have been proposed and investigated. A drawback of this approach is that one cannot easily perform the local unitary transforms such as $\hat{Z}$ operation due to non-orthogonality of two coherent states.

In this approach, the Bell measurement can be implemented as follows. If the four Bell states
\begin{equation}
\label{eq:BellB}
\begin{aligned}
\ket{\Phi^\pm}={\cal N_\pm}(\ket{\alpha}\ket{\alpha}\pm\ket{-\alpha}\ket{-\alpha})\\
\ket{\Psi^\pm}={\cal N_\pm}(\ket{\alpha}\ket{-\alpha}\pm\ket{-\alpha}\ket{\alpha}),
\end{aligned}
\end{equation}
where ${\cal N}_\pm=(2\pm2e^{-4|\alpha|^2})^{-1/2}$, pass through a 50:50 beam splitter (BS), the resulting states are
\begin{equation}
\begin{aligned}
\ket{\alpha}\ket{\alpha}\pm\ket{-\alpha}\ket{-\alpha}&\xrightarrow{\rm
BS}&\big(\ket{\sqrt{2}\alpha}\pm\ket{-\sqrt{2}\alpha}\big)\ket{0},\\
\ket{\alpha}\ket{-\alpha}\pm\ket{-\alpha}\ket{\alpha}&\xrightarrow{\rm
BS}&\ket{0}\big(\ket{\sqrt{2}\alpha}\pm\ket{-\sqrt{2}\alpha}\big).
\end{aligned}
\end{equation}
As $\ket{\sqrt{2}\alpha}+\ket{-\sqrt{2}\alpha}$ and
$\ket{\sqrt{2}\alpha}-\ket{-\sqrt{2}\alpha}$ contain respectively an even and odd number of photons, two parity measurements acting on two output modes discriminate between the four Bell states. Here, the even number state $\ket{\sqrt{2}\alpha}+\ket{-\sqrt{2}\alpha}$ yields the case when no photon is detected in both detectors, due to the nonzero overlap of $\bracket{0}{\pm\sqrt{2}\alpha}=e^{-\alpha^2}$, which is counted as a failure. Assuming equal input probabilities of four Bell states, we can then calculate the average success probability of the Bell measurement as $P_s=1-{\cal N}^2_{+}|\bra{0}(\ket{\sqrt{2}\alpha}+\ket{-\sqrt{2}\alpha)}|^2/2=1-(2\cosh{2\alpha^2})^{-1}$. We can also obtain the average photon number included in the Bell states, by averaging $\langle\hat{n}_1+\hat{n}_2\rangle$ for all four Bell states, resulting in
\begin{align}
\bar{n}=|\alpha|^2\Big(\frac{1-e^{-2\alpha^2}}{1+e^{-4\alpha^2}}+\frac{1+e^{-2\alpha^2}}{1-e^{-4\alpha^2}}\Big),
\end{align}
which we will use later to compare our scheme with respect to the success probability for the average photon number used in the process.

{\em Optical hybrid qubits}-- Recently, a hybrid approach was proposed by combining both approaches for single photons and coherent state qubits \cite{SLee13,SWLEE13}. The logical qubit is constructed based on the orthonormal basis, $\big\{\ket{0_L}=\ket{+}\ket{\alpha},~\ket{1_L}=\ket{-}\ket{-\alpha}\big\}$ where $\ket{\pm}=(\ket{H}\pm\ket{V})/\sqrt{2}$. 
It was shown that this approach enables a near-deterministic quantum teleportation as well as deterministic  $\hat{X}$ and $\hat{Z}$ operations using linear optics. In this scheme, the logical Bell-state measurement  can be performed using two independent Bell-state measurements, the standard Bell measurement ($\rm B_s$) for the single photon part and the coherent state Bell measurement described above for the coherent-state part. The logical Bell measurement can be successfully performed unless both Bell measurements on each parts fail so that its success probability is given as $P_s=1-e^{-2\alpha^2}/2$. As the average number of photons contained in any single Bell state is obtained as $\bar{n} = 2 + 2\alpha^2$, we can rewrite the success probability by $\bar{n}$ as $P_s=1-e^{-\bar{n}+2}/2$. 

\subsection{Comparison with our scheme}

\begin{figure}[t]
\centering
\includegraphics[width=3.0in]{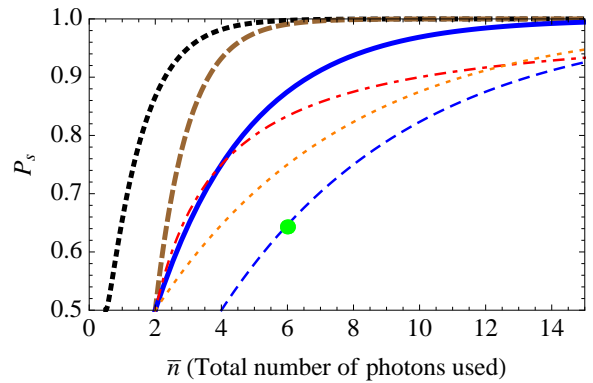}
\caption{(Color online). 
The success probability of Bell measurements against the average photon number ($\bar{n}$) used in the process. The black thick dotted curve is for the coherent state qubits, the brown thick dashed curve for optical hybrid qubits, the blue thick curve for our scheme. The blue dashed curve is for Zaidi and van Loock's scheme2 when $m=2$ (it is the same with ours when $m=1$). The red dot-dashed curve is for Grice's scheme, the orange dotted curve is for Ewert and van Loock's scheme, and the green circle at $P_s=0.643$ and $\bar{n}=6.00029$ for Zaidi and van Loock's scheme1.
}\label{fig:tele}
\end{figure}

In our scheme, the single Bell measurement process consumes a total of $2N$ photons as each logical Bell state contains $2N$ photons and no additional photons are used in the process. Thus, the success probability of Bell measurement can be written as $P_s=1-2^{-\bar{n}/2}$ with the average photon number $\bar{n}=2N$ used in the process. We plot $P_s$ against $\bar{n}$ for our scheme and all other above-mentioned  schemes in Fig.\ref{fig:tele}. We will present some observations about the results and comparisons of our scheme with others as below:

(i) Our scheme shows the best performance among the schemes employing single photon qubits, with respect to the attained success probability against the average photon number ($\bar{n}$) used in the process. For example, our scheme can reaches $P_s=0.996$ by using total $\bar{n}=16$ photons for the Bell measurement. On the other hand, the attained success probabilities by other Bell measurement schemes suggested by Grice \cite{Grice2011}, Zaidi and van Loock \cite{Zaidi2013}, and Ewert and van Loock \cite{Ewert2014} are much lower than ours with the same number of photon usage. Moreover, our scheme does not require photon number resolving detectors in contrast to these schemes  \cite{Grice2011,Zaidi2013,Ewert2014}.

(ii) The scheme recently proposed by Ewert {\em et al.} \cite{Ewert15}, similarly with ours, employs multiphoton entanglement encoding. The logical qubit is constructed in the form of the parity state encoding \cite{Ralph2005}, devised to tolerate photon losses by containing $n$ blocks of $m$ photons, i.e., total $nm$ photons. It also enables one to perform near-deterministic Bell measurement with the success probability, $1-2^{-n}$, similarly with ours. However, in order to achieve $P_s=1-2^{-n}$, it uses total $2nm$ photons. The attained success probability against the average photon number usage is the same as ours when $m=1$. It was pointed out in \cite{Ewert15}  that, under the effects of photon losses, increasing $n$ yields higher success probability than increasing $m$.

(iii) The schemes with coherent state encoding \cite{Jeong2001,Jeong2002} or optical hybrid encoding \cite{SWLEE13,SLee13} yields higher success probabilities than ours for the same average photon number usage as we can see in Fig.\ref{fig:tele}. This is due to the fact that the success probability of coherent state Bell measurement, employed in both (coherent state encoding and hybrid encoding) schemes, dramatically increases as the amplitude of the coherent states gets higher. However, it requires the exact discrimination of even and odd number of photons, which becomes more difficult to realize for higher amplitudes. We stress again that, in contrast, our scheme only requires photon on-off detectors, which is a considerable advantage to realize.

\section{Application to Quantum Information Processing}
\label{sec:App}

In this section, we will consider the application of our scheme to quantum communication and all-optical quantum computation. We will also investigate the effects of photon loss, which is the major detrimental factor of errors for the optical implementation of quantum information processing \cite{Ralph2010}.

In our analysis, we have made some assumptions as follows. The resource states required for logical qubits or entangled channels for gate teleportation, e.g., GHZ entanglement of multiphotons, are assumed to be prepared by off-line processes. The evolution of optical quantum states in a lossy environment is obtained by solving the master equation \cite{Phoenix90},
\begin{equation}
\frac{d\rho}{dt}=\gamma(\hat{J}+\hat{L})\rho,
\end{equation}
where $\hat{J}\rho=\sum_i\hat{a}_i\rho\hat{a}_i^{\dag}$,
$\hat{L}\rho=-\sum_i(\hat{a}^{\dag}_i\hat{a}_i\rho+\rho\hat{a}^{\dag}_i\hat{a}_i)/2$,
and $\hat{a}_i (\hat{a}_i^{\dag})$ is the annihilation (creation)
operator for the $i$-th mode. Here, $\gamma$ is the decay constant, and the loss rate is defined as $\eta=
1-e^{-\gamma t}$. It is assumed that all optical systems, from the logical qubits to teleportation channels, experience the same loss effects with rate $\eta$. During the generation process of entangled quantum channel, loss also occurs with rate $\eta$ so that some imperfect resource states in which photons may be lost at one or more modes are possibly supplied into in-line communication or computation. The logical qubits are assumed to be prepared in ideal pure states when they are first supplied into the in-line process. During the in-line process, the same loss rate $\eta$ is applied to each photonic mode of the qubits for each gate operation and corresponding time in quantum memory. Details of our analysis will be presented in the following subsections.

\subsection{Quantum communication}

Our Bell measurement scheme is useful for the implementation of quantum communications. At first, it immediately enhances the success probability of the standard quantum teleportation \cite{Bennett93}. In our approach, an unknown qubit is prepared with $N$ photon encoding as $\ket{\phi_N}_A=a\ket{+}^{\otimes N}_A+b\ket{-}^{\otimes N}_A$ at site $A$. The channel state $|+\rangle_A^{\otimes N}|+\rangle_B^{\otimes N}+|-\rangle_A^{\otimes N}|-\rangle_B^{\otimes N}$ is distributed between site $A$ and $B$. Here, both the qubit and channel states are in the form of GHZ entanglement. First, the sender located at site A carries out our Bell measurement scheme, {\em i.e.} $N$-times of $\rm B_s$ measurements, where each $\rm B_s$ is acting on two photons, one from $\ket{\phi_N}_A$ and the other from the site $A$ of the channel state. The receiver at site $B$ can then retrieve the input state $\ket{\phi_N}$ by performing appropriate unitary transforms, either Pauli X (bit flip) or Z (phase flip) or both operations, which can be implemented deterministically as we will explain in the following Sec.~\ref{sec:UGO}. The success probability of quantum teleportation with $N$ photon encoding is just given as $P_s=1-2^{-N}$, equal to that of the logical Bell measurement. Therefore, a near-deterministic quantum teleportation can be achieved by employing only linear optics and photon on-off detectors. For example, if four-photon entangled logical qubit and eight-photon entangled channel states are used, the success probability of quantum teleportation becomes higher than 90\%.

In the realization of quantum communications, photon loss is unavoidable. An arbitrarily encoded qubit $\ket{\phi_N}=a\ket{+}^{\otimes N}+b\ket{-}^{\otimes N}$ evolves under a lossy environment into a mixed state in the form of
\begin{equation}
\begin{aligned}
\label{eq:evolved}
&(1-\eta)^N\ket{\phi_N}\bra{\phi_N}+\frac{1}{2}\sum^{N}_{k=1}\binom{N}{k}(1-\eta)^{N-k}\eta^{k}\\
&\qquad\qquad\times\big(\ket{\phi_{N-k}}\bra{\phi_{N-k}}+\ket{\phi^{-}_{N-k}}\bra{\phi^{-}_{N-k}}\big),
\end{aligned}
\end{equation}
where $\binom{N}{k}$ denotes the binomial coefficient and $\eta$ is the loss rate on each photonic mode. As one can see here, the total loss probability of each logical qubit is given by $P=1-(1-\eta)^{N}$. The probability that $k$ photons are lost out of $N$ is $\binom{N}{k}(1-\eta)^{N-k}\eta^{k}$ and its possible resulting state is either $\ket{\phi_{N-k}}=a\ket{+}^{\otimes N-k}+b\ket{-}^{\otimes N-k}$, or $\ket{\phi^{-}_{N-k}}=a\ket{+}^{\otimes N-k}-b\ket{-}^{\otimes N-k}$. Note that there is no logical error in the former state, while the latter contains a Pauli-$Z$ error. Therefore, if a photon is lost from any single mode of a logical qubit, it experiences a Pauli-$Z$ error with probability $1/2$. 

\begin{figure}[t]
\centering
\includegraphics[width=3.0in]{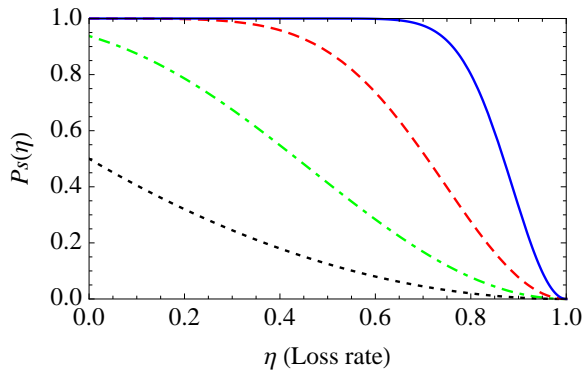}
\caption{(Color online). 
The success probabilities of quantum teleportation under the effect of photon losses ($\eta$) based on our Bell measurement scheme with $N$-photon encoding; the black dotted curve for single photon ($N=1$), the green dotdashed curve for $N=4$, the red dashed curve for $N=16$, and the blue curve for $N=80$ photon encoding.
}\label{fig:succploss}
\end{figure}

The logical Bell measurement is also affected by photon losses. During the quantum teleportation process, losses occur possibly at both the input qubit and the channel states. We thus here consider the Bell measurements performed on two qubits in the assumption that they both experience photon losses with the same rate $\eta$. In our Bell measurement scheme (see Fig.~\ref{fig:scheme}(c)), $N$-times of $\rm B_s$ measurements are performed independently on two photons; one is from the qubit and the other from the channel. If either of the two photons are lost, the $\rm B_s$ fails. The probability that losses occur at $k$-times of $\rm B_s$ can be written by
\begin{equation}
\begin{aligned}
\label{eq:ktimesfail}
&\binom{N}{k}\Big\{(1-\eta)^2\Big\}^{(N-k)}\Big\{1-(1-\eta)^2\Big\}^{k}\\
&=\binom{N}{k}(1-\eta)^{2(N-k)}\eta^{k}(2-\eta)^{k}.
\end{aligned}
\end{equation}
Therefore, we can obtain the total success probability of Bell measurements under losses as
\begin{equation}
\begin{aligned}
\label{eq:succpro2}
P_s(\eta)&=1-\sum^{N}_{k=0}\binom{N}{k}(1-\eta)^{2(N-k)}\eta^{k}(2-\eta)^{k}\left(\frac{1}{2}\right)^{N-k}\\
&=1-\left(\frac{1+\eta(2-\eta)}{2}\right)^N.
\end{aligned}
\end{equation} 
which becomes $P_s(0)=1-2^{-N}$ when there is no loss $\eta=0$ and $P_s(1)=0$ for $\eta=1$. Assuming that no additional operational errors occur during the Bell measurement and feedforward operations, the success probability of teleportation under a lossy environment is obtained as $P_s(\eta)$. We plot $P_s(\eta)$ against the loss rate $\eta$ for different encoding numbers $N$ in Fig.\ref{fig:succploss}. It shows us that in principle quantum teleportation can be implemented with arbitrary high success probability under a lossy environment by increasing the encoding number $N$ in our scheme. Note that here we do not consider the Pauli-Z error caused by losses, which can be propagated to the teleported qubit with probability 1/2. Such a Pauli logical error should be corrected by employing error correction codes, details of which will be considered in the following subsections.

For a long distance quantum communication, the distance between the sender and receiver is limited by the effects of photon losses and quantum repeaters are necessary at intermediated locations \cite{LDQC}. As quantum repeaters typically require Bell measurements to complete the protocol \cite{Ewert15,Azuma15}, our scheme can be possibly emloyed to enhance the performance of such a protocol, which will be a possible future work.

\subsection{Universal gate operations}
\label{sec:UGO}

In order to realize universal quantum computation, a set of logical quantum gates are necessary, by which any logical operation is in principle possible to construct. For example, Pauli X, arbitrary Z (phase) rotation, Hadamard, and a controlled-Z (CZ) operations constitute such a universal set. In our framework, Pauli X and arbitrary Z (phase) operations are straightforward to implement using linear optical elements. The Pauli X (bit flip) operation can be performed by flipping $\ket{+} \leftrightarrow \ket{-}$ implemented by a polarization rotator at all photon modes. An arbitrary Z (phase) operation can be performed by applying the phase shift operation only a single photonic mode: $\{\ket{+},~\ket{-}\} \rightarrow \{\ket{+},~e^{i\theta}\ket{-}\}$, and no operation is required for the other modes. Thus, both the Pauli X and arbitrary Z (phase) operations are deterministic in our approach.

In order to realize the Hadamard and CZ gates, we here employ the gate teleportation protocol 
with specific types of entangled states \cite{Gottesman1999}. For the Hadamard operation, entangled state $\ket{Z}\approx\ket{0_L}\ket{0_L}+\ket{0_L}\ket{1_L}+\ket{1_L}\ket{0_L}-\ket{1_L}\ket{1_L}$ should be used as the quantum channel for teleportation. If one perform a standard teleportation of unknown qubit $\ket{\phi}$ with the channel $\ket{Z}$, the output state is given as the $\hat{H}\ket{\phi}$ where $\hat{H}$ indicates the Hadamard gate. Likewise, one can realize the CZ gate operation by performing two-qubit teleportation with $\ket{Z'}\approx\ket{0000}+\ket{0011}+\ket{1100}-\ket{1111}$ where $\ket{0000}=\ket{0_L}\ket{0_L}\ket{0_L}\ket{0_L}$ and so on, as the quantum channel for teleportation. In this case, the cost is preparation of mutiphoton entanglement as resource states. As the success probability of the Hadamard or CZ gates is the same as the success probability of the logical Bell measurements, it is possible to implement these operations nearly deterministically by increasing the number of photons $N$ in a logical qubit. 

Alternatively, other approaches for the realization of the Hadamard and CZ gates, such as the Knill-Laflamme-Milburn (KLM) scheme \cite{Knill2001} or the parity encoded scheme used in Ref.~\cite{Ralph2005,Gilchrist2005,Hayes2010}, may be considered: Such approaches may possibly reduce the resource requirements in gate operations in comparison to gate teleportation in which many resources are consumed to prepare the channel states $\ket{Z}$ and $\ket{Z'}$ (details will be discussed in Sec.~\ref{sec:FTQC}), however it may require more feedforward operations than gate teleportation.

The effects of photon losses should be also considered for the implementations of quantum computing. A logically encoded qubit with $N$ photons may be affected by losses during the in-line process, resulting in the same form as Eq.~(\ref{eq:evolved}), which contains some logical errors. When performing gate teleportation for the Hadamard and CZ gate operations, the success probability of Bell measurement is also deteriorated by photon losses. Here we assume that the resource channels $\ket{Z}$ and $\ket{Z'}$ are prepared at off-line and immediately used such that only the input qubits are subjected to losses. In this case, the Bell measurement fails with rate $(1/2)^{N-k}$ when $k$ photons are lost in the qubit. Thus, we can calculate the success probability of the gate teleportation with loss rate $\eta$ as
\begin{equation}
\begin{aligned}
\label{eq:master}
P'_s(\eta)&=1-\sum^{N}_{k=0}\binom{N}{k}(1-\eta)^{N-k}\eta^{k}\left(\frac{1}{2}\right)^{N-k}\\
&=1-\left(\frac{1+\eta}{2}\right)^N,
\end{aligned}
\end{equation} 
where $\binom{N}{k}$ represents the binomial coefficient. It becomes $P'_s(0)=1-2^{-N}$ when there is no loss $\eta=0$, and $P'_s(1)=0$ for $\eta=1$. It is higher than the one obtained in Eq.~(\ref{eq:ktimesfail}) as we assumed here that loss occurs only at the input qubit modes (no loss at the channel states).

During the computation process, several errors are caused by photon losses. Some of the errors are detectable by measurements during the gate operations, which are called ``locatable'' \cite{Ralph2010}, while other errors propagates and can only be corrected by an error correction code. This latter kind are called ``unlocatable''. As we can see in Eq.~(\ref{eq:evolved}), losses occur with the rate $P$ in quantum memory or gate operations. If any single photon is lost, the logical qubit experiences a Pauli $Z$ error with probability $1/2$. The loss errors in quantum memory as well as deterministic operations such as the Pauli $X$ or arbitrary $Z$ rotation, are unlocatable. On the other hand, the errors in Hadamard or CZ gates caused by losses are in fact immediately detectable by missing photons at the detectors during the logical Bell measurement. Moreover, as the gate teleportation scheme is used here, missing photons in the input qubit are compensated at the output qubit as far as the teleportation succeeds. 

\subsection{Fault-tolerant quantum computation}
\label{sec:FTQC}

\begin{table}[t]
\begin{tabular}{|c|c||c|c|}
\hline
~$N$~~&~Noise threshold $\eta$~~&~$N$~~&~Noise threshold $\eta$~~\\
\hline
~3~~&~$1.3\times 10^{-3}$~~&~7~~&~$1.1\times 10^{-3}$~~\\
\hline
~4~~&~$1.7\times 10^{-3}$~~&~8~~&~$0.9\times 10^{-3}$~~\\
\hline
~5~~&~$1.5\times 10^{-3}$~~&~9~~&~$0.7\times 10^{-3}$~~\\
\hline
~6~~&~$1.3\times 10^{-3}$~~&~10~&~$0.6\times 10^{-3}$~~\\
 \hline
\end{tabular}
\caption{\label{tab:table1}{Fault-tolerant noise thresholds ($\eta$) for different numbers of photons in a logical qubit ($N$) using the seven-qubit STEANE code and the telecorrection protocol \cite{Dawson2006}. The highest threshold is obtained when $N=4$. }}
\end{table}

In order to build arbitrary large-scale quantum computers, it should be justified that they can be implemented with tolerance to small errors, so-called fault-tolerance \cite{Shor1996}. For this, the amount of noise per operation with appropriate error corrections should be below a fault-tolerant threshold. We thus carried out numerical simulations to obtain the threshold for our scheme with a given loss rate $\eta$. We here employ the seven qubit STEANE code \cite{Steane1996} with several levels of concatenation based on the circuit-based telecorrection \cite{Dawson2006}, in order to directly compare its performance with previous other schemes employing the same correction code and circuit \cite{Dawson2006,Lund2008,Hayes2010,SWLEE13}. Note that the STEANE code can correct arbitrary logical or unlocatable errors, but for the purpose of this calculation we assume that the other errors are negligible compared to the loss errors. 

We will employ the modified telecorrection circuit composed of CZ, Hadamard, $\ket{+}$ state, and $X$-basis measurement \cite{Lund2008}. Let us define the error model for our approach as follows: When Hadamard or CZ gates fail, the output qubit is assumed to experience depolarization, which can be modeled by a random Pauli operation, {\it i.e.} Pauli $Z$ and $X$ operation independently act on the logical qubit with the equal probability $1/2$. For the other loss errors in quantum memory or gate operations, Pauli $Z$ operation is applied to the qubit with probability $1/2$. We use this error model for the lowest level of concatenation, while the same error models described in Ref.~\cite{Dawson2006} is used for higher levels of concatenation. Based on this approach, we performed a series of Monte Carlo simulation (using C++) to obtain the corrected error rates for a range of $\eta$ with different $N$. The resulting corrected error rates in a lower level are used for the error rates for the elements used in next level of error correction. If the error rates tend to zero in the limit of many levels of concatenation, it is guaranteed that in principle the fault-tolerant quantum computation is possible with those certain $\eta$ and $N$. 

In this way, the noise thresholds of our scheme with different $N$ are obtained as shown in Table~\ref{tab:table1}. Interestingly, the largest threshold is obtained when the logical qubit is encoded with four photons ($N=4$). Note that a further increase of $N$ lowers the threshold, which may be due to unlocatable errors becoming dominant for larger $N$ encoding. Let us then compare the result with those obtained by other schemes. The threshold obtained by our scheme ($\sim 1.7\times10^{-3}$) is much higher than those for coherent-state qubits ($\sim2\times10^{-4}$) \cite{Jeong2001,Jeong2002,Ralph2003,Lund2008} and optical hybrid qubits ($\sim5\times10^{-4}$) \cite{SWLEE13}. On the other hand, it is almost equivalent to the one using parity states encoding with entangled photons \cite{Ralph2005,Hayes2010}. Even much higher thresholds may be attainable by employing recently proposed topological error codes \cite{Yao2012Nature,Sean2010}, which will be a possible future work.

The number of resources required for one round of error correction is another important factor to analyze the scalability of quantum computation. One can estimate the resource requirements of the telecorrection scheme by the following estimation. The total number of operations in one round of telecorrection is about 1,000 \cite{Ralph2010} and the element operations are used with the following fractions \cite{Dawson2006,Lund2008}: memory 0.284, Hadamard 0.098, CZ 0.343, Diagonal state 0.164, and X-basis measurement 0.111. If we assume that the number $N_{H}$, $N_{\rm CZ}$, and $N_{\ket{+}}$ of resources are consumed to prepare the Hadamard, CZ gate, and the diagonal state, respectively, and none for the others. Then, the total resource requirements for one round of error corrections is obtained by
\begin{equation}
\nonumber
98N_{H}+343N_{\rm CZ}+164N_{\ket{+}}.
\end{equation}
In fact, it depends on the form of the required entangled states and the efficiency of their generation methods. In our scheme, as explained in Sec.~\ref{sec:UGO}, the entangled states $\ket{Z}$ and $\ket{Z'}$ are used as the channel for gate teleportation to realize Hadamard and CZ gate, respectively. 
If we assume conditional generation of $\ket{Z}$ and $\ket{Z'}$ from entangled photon pairs {\em e.g.} using the fusion gate operation \cite{DEBrown2005}, the resource cost increases exponentially with increasing $N$ the number of photon in a logical qubit.
On the other hand, any on-demand or efficient generation scheme for such multiphoton entanglement will be able to reduce dramatically the resource requirements. 
Recently, such multiphoton entanglement has been experimentally produced \cite{JWPan2012}. For example, GHZ entanglement up to eight photons \cite{Yao2012,Huang2011} and cluster states up to eight photons \cite{Yao2012Nature} have been successfully generated. Several on-demand generation schemes were proposed theoretically \cite{deterministic,Lindener2009} and are expected to be realized, e.g., based on semiconductor quantum dots \cite{Young2006}. 

\section{Conclusions}

We have presented a detailed scheme for nearly deterministic Bell measurement using multiphoton entanglement and linear optics proposed in Ref.~\cite{SLee15}. The limitation that only two of four Bell states can be identified by the standard Bell measurement, $\rm B_s$, is overcome by the GHZ type of multiphoton entanglement and $N$ times of $\rm B_s$ measurements, where $N$ is the number of photons in a logical qubit. The logical Bell measurement is performed through $N$ times of $\rm B_s$ measurements, and it fails only when none of those $N$ times of $\rm B_s$ measurements succeeds. Therefore, the success probability of the logical Bell measurement is given as $1-2^{-N}$, which rapidly approaches to unity as $N$ increases.

We have compared the efficiency of our scheme with those of other previous schemes proposed to improve the success probabilities of Bell measurement based on linear optics; Several schemes were proposed based on single photon qubits \cite{Grice2011,Zaidi2013,Ewert2014,Ewert15} or other optical qubits such as coherent state \cite{Jeong2001,Jeong2002} and optical hybrid qubits \cite{SWLEE13}. We showed that our scheme outperforms other schemes employing single photon qubits, with respect to the attained success probability in terms of the average number of photons used in the process. Although the scheme using coherent state or optical hybrid qubits achieves higher success probability than ours by consuming the same number of photons on average, a considerable advantage of our scheme over all the other schemes is that it does not require photon number resolving detections but only photon on-off detection suffices.

We have demonstrated that our scheme can be effectively used for the implementation of quantum teleportation and all-optical quantum computation. The success probability of quantum teleportation increases rapidly to unity as increasing $N$ the number of photons in the qubit and channel states. We showed that in our approach fault-tolerant quantum computation is possible under photon losses. It is remarkable that the highest noise-threshold is obtained with four-photon encoded qubits and eight-photon entangled channels that are accessible within current technologies \cite{JWPan2012}. For example, 8 photon GHZ entangled states have been recently generated successfully in optics laboratories \cite{Yao2012,Huang2011}.

Interestingly, our Bell measurement scheme can be implemented via either spatially or temporally distributed and independent $N$ times $\rm B_s$ measurements. Such an experiment may be considered not only with spatially separated $N$-mode entangled states but also with temporal mode entanglement as done in Refs.~\cite{ex1,ex2,ex3}. For a temporally distributed implementation, only one standard Bell-measurement device illustrated in Fig.\ref{fig:scheme}(a) \cite{Lut99} will be sufficient to perform independent $N$ times of $\rm B_s$ measurements for a logical Bell measurement. 
We finally note that our idea for a distributed Bell measurement scheme, in principle, is not limited to photonic qubits but can be applied to any multipartite systems with other physical degrees of freedom. Our works reveals the possibility of using multipartite entanglement for efficient quantum communication and computation.

\section*{Acknowledgements}
We would like to note that the comparison in our paper with other schemes \cite{Grice2011,Ewert2014,Zaidi2013,Ewert15} does not rule out their own advantages for the implementation of Bell measurement with linear optics. We thank F. Ewert and P. van Loock for valuable comments and discussions.
This work was supported by the National Research Foundation of Korea (NRF) grant funded by the Korea government (MSIP) (Grant No. 2010-0018295). K.P. acknowledges financial support from Grant No. GA14-36681G of the Czech Science Foundation.
T.C.R. was supported by the Australian Research Council Centre of Excellence for Quantum Computation and Communication Technology (Project No. CE110001027).

\end{document}